# ChatGPT-5 in Secondary Education: A Mixed-Methods Analysis of Student Attitudes, AI Anxiety, and Hallucination-Aware Use

Tryfon Sivenas

National and Kapodistrian University of Athens

School of Pedagogical and Technological Education

sivenastrif@primedu.uoa.gr

**Abstract**

This study employed a mixed-methods design in order to explore secondary students' interactions with the generative AI chatbot ChatGPT-5 in a formal classroom setting, focusing on attitudes, anxiety and on responses to hallucinated outputs. Participants were 109 sixteen-year-old students from three Greek high schools who engaged with ChatGPT-5 during an eight-hour intervention integrated into the course "Technology." Students interacted with the chatbot through multiple modalities such as information seeking, CV generation, document and video summarization, image generation, student-generated quizzes, and custom age-related explanations, including tasks deliberately designed to trigger hallucinated responses. Quantitative data was collected with the Student Attitudes Toward Artificial Intelligence scale (SATAI) and the Artificial Intelligence Anxiety Scale (AIAS), and qualitative data was obtained from semi-structured interviews with 36 students. SATAI results indicated moderately positive attitudes toward AI, with more favorable cognitive evaluations than behavioral intentions. AIAS findings revealed moderate learning-related anxiety and higher concern about AI-driven job replacement. Gender differences were small and non-significant for AI anxiety, whereas female students reported significantly more positive cognitive attitudes toward AI than males. AI attitudes and AI anxiety were essentially uncorrelated. Qualitative analysis identified four pedagogical affordances (knowledge expansion, immediate feedback, familiar interface, and perceived skill development) and three constraints (uncertainty about accuracy, anxiety about AI feedback, and privacy concerns). After encountering hallucinations, many students reported restricting AI use to domains where they already had knowledge and could verify answers, a strategy termed epistemic safeguarding. The study concludes with implications for critical AI literacy in secondary education.

## 1. Introduction

Advances in Machine Learning (ML), Deep Learning (DL) and Natural Language Processing (NLP) have rapidly accelerated the development of Artificial Intelligence (AI) (Klar, 2025). In research literature there are several attempts at defining AI, however a recent one from Gignac & Szodorai (2024) suggests that AI is a field of computer science that deals in creating systems capable of solving complex tasks that require perception, prediction and decision-making. Within the broader field of computational intelligence, Generative AI (Hereinafter referred to as GenAI) refers to models that learn statistical patterns from a number of large datasets and in return generate novel outputs in response to natural-language prompts (Wang & Fan, 2025). These may include among others text, images, video, audio and code. A few examples of a prominent subclass are Large Language Models (LLMs), including among others ChatGPT, Gemini, Claude, CoPilot and LLaMA. These systems are capable of interpreting and generating human-like text thus supporting a wide range of tasks including, explanation and summarization to programming and reasoning (Wang & Fan, 2025). Large Language Models (LLMs) operate at the intersection of computational intelligence and Human–Computer Interaction (HCI) and are increasingly applied across diverse domains, including education (Yang et al., 2025).

Education has emerged as one of the earliest and most visible domains of early GenAI adoption (Chiu, 2023). Earlier research in education, however, largely occurred under the umbrella of Artificial Intelligence in Education (AIED). That body of work focused among others on intelligent tutoring systems, adaptive learning environments, as well as educational data mining (Ng et al., 2021). Many of these systems are grounded in classical computational intelligence methods such as fuzzy logic, evolutionary search and neural-network-based student modelling, which are used to infer learners' states and adapt instructional content accordingly. In this sense, GenAI chatbots such as ChatGPT-5 can be seen as the latest layer in a longer trajectory of computational-intelligence-driven tools that mediate between learner input, system inference and pedagogical feedback.

Since the emergence of LLM-based chatbots in late 2022 there has been a surge of publications exploring how GenAI can support teaching and learning (Alfarwan, 2025; Wang & Fan, 2025). Recent literature, scoping, systematic reviews and meta-analyses present mixed findings. Even though GenAI can enhance learning, performance and student active engagement (Deng et al., 2024), it also raises concerns about bias, academic integrity, over-reliance and overall erosion of critical thinking (Lo et al., 2024; Yang et al., 2025).

A limitation of GenAI is hallucination, which refers to the production of confident and plausible but incorrect responses, inconsistent or unsupported by data (Zhang et al., 2025). In educational contexts, hallucinations may appear as fabricated and/or incorrect explanations that include fictitious references which in turn can mislead learners (Valeri et al., 2024) particularly those with limited prior knowledge to evaluate the generated content (Ding et al., 2023). Although existing studies on GenAI in education identify hallucination as a concern (Long & Magerko, 2020; Ng et al., 2021), few have intentionally designed activities that expose students to hallucinated content.

Recent mapping and systematic reviews (Deng et al., 2024) highlight that GenAI research has concentrated primarily on university education (Lo et al., 2024) and to a lesser extent on primary and secondary school contexts (Yang et al., 2025). A recent systematic review (Alfarwan, 2025) noted that the majority of studies are of small-scale, short-term, and predominantly concentrated in STEM subjects across both in formal and informal learning environments. Similarly, a scoping review (Yim & Su, 2024) reported similar patterns and highlighted a geographical concentration of studies in East and Southeast Asia as well as United States. Overall, the majority of GenAI related interventions occur in technologically advanced regions whilst other regions are under-represented (Ng et al., 2021; Zhang et al., 2025). Another recurring limitation in literature is that although most studies specify the model used (e.g., ChatGPT), they provide limited information regarding the intervention itself or the modalities through which students interacted with the model (e.g., text-to-text, document-to-text, video-to-text, text-to-image) (Deng et al., 2024). Detailed reporting on GenAI interaction modalities would help provide a clearer understanding of students' needs and help clarify whether the model is used mainly as an answer-retrieval tool or as a means of supporting deeper conceptual engagement that provides self-regulation and critical evaluation.

Therefore, the purpose of this study is to address the gaps identified above, by examining Greek high school students' responses to a GenAI supported intervention using ChatGPT-5 within the national curriculum course "Technology". The intervention comprised seven clearly defined activities (information seeking, customized Curriculum Vitae generation, document and video summarization, image generation, student generated quizzes on topics of interest and custom age-related explanations). Students were also exposed to examples of hallucinated outputs, which were discussed collectively. The study adopts a mixed-method design that integrates validated instruments for measuring Student Attitudes Towards AI (SATAI) as well as AI-related anxiety (AIAS) with qualitative analysis of students' affordances and constraints when using GenAI in educational settings.

Additionally, this study is guided by the following research questions:

RQ1: What are Greek secondary students' attitudes toward GenAI?

RQ2: What levels of GenAI-related anxiety do Greek secondary students report?

RQ3: What affordances and constraints do students perceive when using GenAI for learning?

RQ4: How do students navigate concerns related to GenAI accuracy and hallucinations?

The remainder of the paper is structured as follows: Section 2 presents a review of the literature, Section 3 describes the methodology, Section 4 presents the quantitative and qualitative results. Section 5 discusses the results lastly, Section 6 outlines limitations and directions for future work.

## 2. Literature review

Four research databases were searched (IEEEXPLORE, Scopus, SpringerLink, ERIC) between September and November 2025, following PRISMA guidelines (Page et al., 2021). The search strategy used a combination of the following keywords: "Generative AI OR Large Language Models OR ChatGPT OR Gemini OR Claude AND Secondary education OR K-12 Or school students AND perceptions OR attitudes OR AI literacy OR AI anxiety OR hallucinations OR affordances OR constraints". The search was limited to studies published between 2020 and 2025. The initial search produced 1481 records, which were screened for relevance based on titles, abstract and alignment with scope of the present study. Four inclusion criteria were applied (a) empirical studies, (b) published in English, (c) focusing on secondary education and (d) aligned with the aims of the review. Studies were excluded if they (a) were purely technical, theoretical, reflective or opinion-based without empirical data, (b) focused on educational levels other than secondary (e.g., university education), or (c) did not report original research findings. After removing duplicates, 29 studies remained. A close examination of the inclusion and exclusion criteria reduced the number to nine eligible papers.

Alneyadi and Wardat (2023) conducted a mixed-method quasi-experimental study examining the impact of ChatGPT on learning electromagnetism. The sample consisted of 122 eleventh-grade students from secondary schools in the United Arab Emirates, divided into two groups: control (N=64) and experimental (N=58). Moreover, the experimental group was encouraged to use ChatGPT as a learning assistant whenever they needed support. Data was collected through (a) pre/post testing, (b) open-ended surveys as well as semi-structured interviews. After four weeks of using ChatGPT, the experimental group demonstrated substantially higher post-test performance with improved understanding of several complex physics concepts. Both male and female students reported positive perceptions, although female students indicated more frequent use of ChatGPT than their male peers. The authors concluded that ChatGPT can enhance achievement and perceived learning in secondary education.

On a different approach Bitzenbauer (2023) conducted a pilot study with 53 twelve grade students (aged 17-18 years old) in Germany to examine their perception of using ChatGPT-3 in physics learning. The intervention comprised two 45-minute sessions in which students generated and evaluated information regarding several physics concepts (e.g., photons, wave-particle duality). Data was collected via pre/post questionnaires targeting students perceptions of AI. Results indicated that the majority of students showed a positive shift recognizing ChatGPT's usefulness, benefits and future relevance. The author concluded that interacting with ChatGPT fostered critical thinking, increased students' optimism and enhanced their awareness of the model's capabilities.

Chen and Chang (2024) conducted a quasi-experimental study investigating how ChatGPT could support 202 seventh-grade students in Taiwan during a digital game-based learning activity on Newtonian mechanics. The intervention introduced the educational physics game "Summon of Magicrystal". Students were assigned into three groups (a) game only control group, (b) ChatGPT-assisted group, and (c) ChatGPT-assisted group with additional prompting examples. Data was collected through pre/post testing, motivation and cognitive load questionnaires, game log analysis and lastly interviews. Results indicated that both ChatGPT supported groups outperformed the control group in learning gains and experiences lower cognitive load.

Ng et al. (2024) conducted a mixed-method study comparing two chatbots, (a) a ChatGPT-based chatbot (SRLbot) and a rule-based chatbot (Nemobot) in order to examine self-regulated learning (SRL) among 74 secondary students in Hong Kong. The intervention lasted 3 weeks (10 lessons in total) and data was collected via pre/post testing using the Online SLR questionnaire (Zimmerman & Schunk, 2011) and Chinese Motivated Strategies for Learning Questionnaire (Ng & Chu, 2021) as well as semi-structured interviews. Results indicated that the group that used SRLbot demonstrated greater improvements across all SRL dimensions. In contrast, Baha et al. (2023) examined the use of educational chatbots in order to support learning of the Logo programming language in 109 secondary school students from Morocco. The intervention was split into five two-hour sessions where the students interacted with the chatbot. Data was collected using pre/post tests that inquired about logo knowledge, classroom activities as well as an overall satisfaction questionnaire. Results indicated that students who used the chatbot reported increased motivation and reduced stress.

In a more recent work, Klar (2025) examined 106 German secondary school students' perceptions and interaction patterns when using GenAI chatbots in a 20-minute research task about cognitive biases. The students were split into control and experimental group. Data was collected using pre/post surveys (targeting cognitive load, self-regulated learning skills and satisfaction), open-ended tests and a qualitative analysis of the chat logs. Results indicated no significant differences in knowledge gain or cognitive load in both groups.

On a different approach, Zhang et al. (2025) conducted a survey including 622 secondary school students in China that examined AI learning attitude, literacy as well as career interest. Notably, students in this study did not seem to interact with GenAI nor an intervention was implemented. Data was collected using the AI learning Attitude Survey (Shah & Mahmood, 2011), the self-developed AI literacy Survey, as well as the AI career interest survey (adapted from Kier et al., 2013). Results indicated that students' learning attitudes positively predicted both AI literacy and AI-related career interest. An additional finding related to gender, showed that female students scored lower on all three constructs.

Yang et al. (2025) investigated the effectiveness of ChatGPT as a virtual teaching assistant for 153 high school students in Taiwan during a six-week C++ programming course. After three weeks of training the students were split into control and experimental that used ChatGPT for three weeks. Data was collected using programming achievement test, flow experience scale (Pearce et al., 2005), Programming self-efficacy scale (Wang & Lin, 2006) as well as post-intervention interviews. Results showed that the experimental group exhibited lower achievement and self-efficacy which the authors attributed to unclear explanations, errors (hallucination) and students' overreliance to ChatGPT.

Lastly, Wu and Zhang (2025), surveyed 500 students on the matter of how GenAI related to the ability to innovate and examined digital literacy as well. The authors measured the students' daily interactions of ChatGPT using the AI application scale and Innovation Capability and Digital Literacy scale. Results indicated that broader and deeper self-reported use of GenAI was associated with higher levels of innovation and digital literacy.

Across the identified empirical studies, several patterns emerge. First, sample sizes in secondary education interventions range widely from 53 up to 622 participants, with the findings being consistent regarding the potential and shortcomings of using GenAI in educational settings. Another finding deals with a concentration in STEM related subjects, including Physics (Alneyadi & Wardat, 2023; Bitzenbauer, 2023; Chen & Chang, 2024) and Programming (Baha et al., 2023; Yang et al., 2025). It seems that GenAI-assisted instruction provides improved conceptual understanding, increased motivation and in some instances enhanced self-regulated learning skills were reported (Alneyadi &Wardat, 2023; Ng et al., 2024). Several studies also examined broader constructs such as digital literacy, innovation capability and AI learning attitudes. One example is the work of Wu and Zhang (2025) that demonstrated higher digital literacy and creativity when using GenAI whereas Zhang et al. (2025) demonstrated how AI literacy mediates the relationship between learning attitudes and AI career interest.

Not all studies, however, report positive outcomes. In their work Yang et al. (2025) reported lower achievement and reduced self-efficacy when using GenAI as a teaching assistant. Similarly, Klar (2025) showed that GenAI usage does not automatically translate to student cognitive gains. These findings highlight how inaccuracies, conceptual gaps and hallucinations can pose major challenges for students, especially when they lack sufficient prior knowledge or evaluative skills.

Similar findings are reported in recent systematic reviews around GenAI in educational contexts. For instance, in his work Alfarwan (2025) noted that empirical research in K-12 remains underdeveloped, with the majority of intervention being small-scale, short in duration and heavily STEM focused. Similarly, Yim and Su (2024) commented on how the majority of GenAI research is targeted in university education. Another finding that was mentioned deals with GenAI inaccuracies and inconsistencies, that has not been examined thoroughly in educational contexts.

Therefore, this study addresses these gaps by combining validated instruments (SATAI, AIAS) with a structured intervention that explicitly integrates and examines Greek high school student responses to AI hallucinations.

## 3. Methodology

The present study took place between September and October 2025, in three high schools in the rural area of metropolitan Athens in Attica, Greece. The study adopted a mixed-methods approach, similar to those used in educational technology research. The remainder section of methodology is split into five subsections. Subsection 3.1 deals with participants, 3.2 discusses quantitative data collection via questionnaires, 3.3 tackles qualitative data collection via semi-structured interviews, 3.4 discusses the procedure and lastly 3.5 deals with the analysis section.

### 3.1 Participants

A total of 109 secondary-grade high school students participated voluntarily in the study. All participants were in their second year of study in high school and were 16 years of age. Of these, 65 (59.6%) were female and 44 (40.4%) were male. In total, 34 students (31.2%) reported prior use of GenAI, with all such reports referring specifically to ChatGPT.

### 3.2 Quantitative data collection

Quantitative data was collected through an online questionnaire administered via Google Forms, which consisted of two sections. Section A asked students to provide anonymized demographic information (e.g., gender, age) and to report their prior use of GenAI tools. Section B requested information regarding their attitudes as well as potential anxiety regarding the use of AI in education. The attitudes were measured with the Student Attitude Towards Artificial Intelligence (SATAI) (Suh & Ahn, 2022) which includes 26 items rated on a 5-point Likert scale. The SATAI consists of three dimensions: (a) Cognitive, (b) Affective, and (c) Behavioral Attitudes. AI-related anxiety was assessed using items from the AI anxiety (AIAS) (Wang & Wang, 2019). The AIAS contains 21 items, rated on a 7-point Likert scale that assess four dimensions namely (a) Learning, (b) AI Configuration, (c) Job Replacement, (d) Sociotechnical Blindness. This study made use of two dimensions of AIAS, namely Learning (eight items) and Job Replacement (three items) as they were deemed most relevant to the educational and future-career context of high school students. As both instruments were originally developed in English, a five-step back-translation procedure (Danielsen et al., 2015) was adopted to ensure accurate translation of the items into Greek.

### 3.3 Qualitative data collection

Qualitative data was collected through semi-structured interviews lasting 8-10 minutes. During the interviews, the students responded to questions regarding (a) affordances, (b) constraints, as well as (c) intended future use in both educational and non-educational settings.

### 3.4 Procedure

The intervention was delivered across three stages within the course "Technology", which forms part of the Greek national curriculum, and took place over three days. Moreover, it deals among others with the introduction of emerging technologies in education and with the end goal of creating a physical technological model at the end of the year. In the first stage, which focused on orientation, the researcher delivered a 60-minute introductory presentation on GenAI, including a brief historical overview, recent developments and everyday applications. The second stage consisted of two extended sessions of 180 minutes each. The first session focused on familiarizing students with ChatGPT's 5 interface. Moreover, the students created individual accounts where required and were encouraged to explore ChatGPT-5 freely.

During the second 180-minute session the students had to use four GenAI interaction modalities, namely text-to-text, text-to-image, document-to-text as well as video-to-text. Moreover, they were tasked to:

1. Seek information on a personal topic of interest. The purpose of this task was to familiarize students with asking questions and following up (prompt engineering).
2. Create their own Curriculum Vitae for a hypothetical summer job. There, the students were presented with a scenario where they were planning on getting a two-month summer job. They were asked to provide personal information, including hobbies and activities, in order to get a personalized resume aligned to their hobbies. The generated CV was saved as a Microsoft Word file.
3. Summarize a document. The students were tasked with locating an article/document of interest and they would then upload it and ask for a summary and do follow up questions on it.
4. Summarize a video. The researcher asked the students to pick a video of their interest. They were then asked to provide the link to ChatGPT-5 and request a summary of it and do follow up questions around it.
5. Generate an image. Students wrote prompts describing an item of interest, which DALL-E used to generate corresponding images.
6. Student-generated tests. Students selected a school subject they found challenging and asked ChatGPT-5 to generate a five-question quiz on the topic.
7. Custom age-related response. In the last task, students selected a concept they personally found difficult to understand in a school subject (e.g., quadratic equations in mathematics) and asked ChatGPT-5 to explain it for different age groups. For instance, they were instructed to request: "Explain quadratic equations as if you are speaking to a 7-year-old, now to a 10-year-old, and now to a 14-year-old." This activity was designed to demonstrate how GenAI can adapt explanations to different developmental levels and to prompt students to reflect on the clarity and appropriateness of AI-generated explanations.

A central component of the second stage involved giving students direct exposure to AI-generated hallucinations. Firstly, the researcher showed students pre-identified examples of ChatGPT-5 producing hallucinated responses on specific fact-based questions (inventing scientific facts and providing incorrect historical and geographical details) during the presentation. Secondly, the students were instructed to inquire ChatGPT-5 about the same pre-validated to answer incorrectly questions. Then the students had to look up the correct answer. Finally, students were asked to apply this awareness by attempting to identify potential hallucinations on a topic of their choosing.

The third and last stage lasted for 65 minutes. Firstly, the students had 25 minutes to use ChatGPT-5. Then they had 25 minutes in order to complete the online questionnaire. 10 minutes for voluntary interviews where 36 students participated and lastly where five minutes for questions around the intervention.

### 3.5 Analysis

Quantitative analyses were conducted using IBM SPSS Statistics (version 29). Moreover, descriptive statistics including means, standard deviations, minimum and maximum values were calculated for all of the items of SATAI as well as AIAS scales. Additionally, internal consistency reliability was examined using Cronbach's alpha.
Qualitative responses from the semi-structured interviews of 36 students were analyzed using open, axial and selective coding, following the methodology of Corbin and Strauss (2008). This coding approach has been used in emerging educational technologies including drones (e.g., Sivenas & Koutromanos, 2022a) and smart glasses (Koutromanos et al., 2023). Themes were refined to capture students' perceived affordances and constraints on the use of ChatGPT-5 during the intervention.

# 4 Results

Section 4.1 presents the quantitative results of SATAI and AIAS, 4.2 discusses group differences and relationships and lastly, section 4.3 presents the qualitative themes that emerged from the interviews.

## 4.1 Quantitative Results

### 4.1.1 SATAI Scale

The SATAI scale exhibited excellent internal consistency, with Cronbach's alpha of 0.95 for the total score as shown in Table 2. All three subscales showed similarly high reliability, with Cognitive demonstrating (α=0.80), Affective (α=0.90) and Behavioral (α=0.94). Item means ranged from 2.58 to 3.65 (Table 1), suggesting moderately positive attitudes toward the use of GenAI in educational settings.

The highest rated items clustered around the perceived value and usefulness of GenAI. In particular, students agreed that "AI makes people's lives more convenient" (Item 6, M = 3.65, SD = 0.98) and that "it is important to learn about AI in school" (Item 1, M = 3.52, SD = 0.99). Similarly, "AI helps me solve problems in real life" (Item 9, M = 3.47, SD = 1.19) and "lessons about AI should be taught in school" (Item 3, M = 3.46, SD = 1.02).

Table 1. SATAI Descriptive statistics (N=109)

| Item No | Item Description | Subscale | Mean | Std. Deviation |
|---|---|---|---|---|
| 1 | I think that it is important to learn about AI in school | Cognitive | 3.52 | 0.99 |
| 2 | AI class is important | Cognitive | 3.4 | 0.99 |
| 3 | I think that lessons about AI should be taught in school | Cognitive | 3.46 | 1.02 |
| 4 | I think every student should learn about AI in school | Cognitive | 3.38 | 0.92 |
| 5 | AI is very important for developing society | Affective | 3.45 | 1.01 |
| 6 | I think AI makes people's lives more convenient | Affective | 3.65 | 0.98 |
| 7 | AI is related to my life | Affective | 3.3 | 1.2 |
| 8 | I will use AI to solve problems in daily life | Affective | 3.4 | 1.2 |
| 9 | AI helps me solve problems in real life | Affective | 3.47 | 1.19 |
| 10 | I will need AI in my life in the future | Affective | 3.36 | 1.05 |
| 11 | AI is necessary for everyone | Affective | 3.39 | 1.03 |
| 12 | AI produces more good than bad | Affective | 3.25 | 1.04 |
| 13 | AI is worth studying | Affective | 3.38 | 1.14 |
| 14 | I think that most jobs in the future will require knowledge related to AI | Affective | 3.15 | 1.11 |
| 15 | I want to work in the field of AI | Behavioral | 3.07 | 0.94 |
| 16 | I will choose a job in the field of AI | Behavioral | 2.58 | 1.38 |
| 17 | I would participate in a club related to AI if there was one | Behavioral | 2.81 | 1.24 |
| 18 | I like using objects related to AI | Behavioral | 3.16 | 1.24 |
| 19 | It is fun to learn about AI | Behavioral | 3.19 | 1.21 |
| 20 | I want to continue learning about AI | Behavioral | 3.34 | 1.33 |
| 21 | I'm interested in AI-related TV programs or online videos | Behavioral | 2.69 | 1.48 |
| 22 | I want to make something that makes human life more convenient through AI | Behavioral | 2.88 | 1.5 |
| 23 | I am interested in the development of AI | Behavioral | 3.09 | 1.48 |
| 24 | It is interesting to use AI | Behavioral | 3.22 | 1.36 |
| 25 | I think that there should be more class time devoted to AI in school | Behavioral | 3.12 | 1.32 |
| 26 | I think I can handle AI well | Behavioral | 3.37 | 1.24 |

On the other hand, the lower rated items were concentrated in the behavioral dimension. Items such as “I will choose a job in the field of AI” (Item 16, M = 2.58, SD = 1.38), “I’m interested in AI related TV programs or online videos” (Item 21, M = 2.69, SD = 1.48) and “I want to make something that makes human life more convenient through AI” (Item 22, M = 2.88, SD = 1.50) received the lowest mean scores.

Taken together, these findings indicate moderately positive attitudes toward GenAI: students see AI as important and socially valuable (higher cognitive and affective scores: M = 3.44 and M = 3.38 respectively) but show more cautious behavioral intentions (Behavioral M = 3.04), especially when it comes to GenAI related careers or extracurricular engagement (Table 2).

Table 2. SATAI subscale Reliability and Descriptive Statistics

| **Subscale** | **Number of Items** | **Average Mean** | **Average Std. Deviation** | **Cronbach's Alpha** |
|---|---|---|---|---|
| Cognitive | 4 | 3.44 | 0.71 | 0.80 |
| Affective | 10 | 3.38 | 0.80 | 0.90 |
| Behavioral | 12 | 3.04 | 1.01 | 0.94 |
| Total SATAI | 26 | 3.24 | 0.79 | 0.95 |

#### 4.1.2 AIAS Results

The AIAS scores also demonstrated high internal consistency in this sample (overall α = 0.91) (Table 4). The learning anxiety subscale showed excellent reliability (α = 0.91), and the Job Replacement subscale demonstrated good reliability (α = 0.87). Across the 11 items used in this study, means ranged from 2.79 to 4.46 (Table 3), indicating low to moderate learning anxiety and somewhat higher concerns about AI driven job replacement.

Within the learning dimension, item means clustered slightly below the neutral midpoint of the 7-point scale. For example, students reported moderate unease with “Taking a course on how AI technologies or products are developed makes me feel anxious” (L1, M = 3.26, SD = 1.76) and “I feel anxious when I have to learn how to use AI technologies/products” (L2, M =

3.27, SD = 1.65). Items related to reading manuals or keeping up with advances in AI received slightly lower means, such as “Learning how to interact with an AI technology/product makes me anxious” (L6, M = 2.79, SD = 1.74) and “I feel anxious about not being able to keep up with advances in AI technologies/products” (L7, M = 2.93, SD = 1.70). Overall, learning related AI anxiety was modest (Learning M = 3.09, SD = 1.37).

Table 3 AIAS Descriptive statistics (N=109)

| Item No | Item Description | Subscale | Mean | Std. Deviation |
|---|---|---|---|---|
| L1 | Taking a course on how AI technologies or products are developed makes me feel anxious. | Learning | 3.26 | 1.76 |
| L2 | I feel anxious when I have to learn how to use AI technologies/products. | Learning | 3.27 | 1.65 |
| L3 | Learning to understand all the special features of an AI technology/product makes me anxious. | Learning | 3.21 | 1.67 |
| L4 | I feel anxious when learning how an AI technology/product actually works. | Learning | 3.19 | 1.69 |
| L5 | Learning to use particular functions of an AI technology/product makes me anxious. | Learning | 3.01 | 1.57 |
| L6 | Learning how to interact with an AI technology/product makes me anxious. | Learning | 2.79 | 1.74 |
| L7 | I feel anxious about not being able to keep up with advances in AI technologies/products. | Learning | 2.93 | 1.7 |
| L8 | Reading the manual for an AI technology/product makes me anxious. | Learning | 3.44 | 1.98 |
| J1 | I am afraid that an AI technique/product may make us dependent. | Job Replacement | 4.03 | 2 |
| J2 | I am afraid that an AI technique/product | Job Replacement | 4.46 | 1.99 |
| | may make us even lazier. | | | |
| J3 | I am afraid that an AI technique/product may replace humans. | Job Replacement | 4.37 | 1.9 |

In contrast, the Job Replacement items were clearly higher. Students expressed concern that "I am afraid that an AI technique/product may make us dependent" (M = 4.03, SD = 2.00) and especially that "I am afraid that an AI technique/product may make us even lazier." (M = 4.46, SD = 1.99). As a result, the Job Replacement subscale mean (M = 4.07, SD = 1.66) was noticeably higher than the learning subscale.

Table 4 AIAS subscale reliability and descriptive statistics (N=109)

| **Subscale** | **Number of Items** | **Average Mean** | **Average Std. Deviation** | **Cronbach's Alpha** |
|---|---|---|---|---|
| Learning anxiety | 8 | 3.09 | 1.37 | 0.91 |
| Job Replacement anxiety | 3 | 4.07 | 1.66 | 0.87 |

### 4.2 Group differences and relationships

To explore potential group differences, independent samples t tests were conducted for gender and prior ChatGPT use on both scales (SATAI and AIAS). In addition, correlations were computed between attitudes toward GenAI (SATAI) and AI anxiety (AIAS).

Firstly, no significant gender differences emerged from the AIAS scores. Moreover, male students reported slightly higher AI anxiety than female students, both for the overall AIAS score ($M_{male} = 3.58$, $M_{female} = 3.36$; $t(87.45) = 0.82$, $p = .416$, $d = 0.16$) and for the learning and Job Replacement subscales, but these differences were small and non-significant (Table 5). On the matter of SATAI, gender differences were more pronounced. Female students scored higher on the total SATAI score (M = 3.35) than male students (M = 3.06), yielding a small to medium effect ($t(101.12) = -1.99$, $p = .049$, $d = -0.38$). The most robust difference emerged on the cognitive subscale: girls reported substantially more positive cognitive attitudes toward AI (M = 3.68) than boys (M = 3.09; $t(95.23) = -4.71$, $p < .001$, $d = -0.91$, partial $\eta^2 = .19$). Affective and behavioral scores were also higher among female students, but these differences did not reach conventional significance (Table 5).

Secondly, on the matter of prior ChatGPT use, it did not seem to affect AIAS significantly. However, clear differences were observed on GenAI attitudes (SATAI). Students who had previously used ChatGPT reported more positive attitudes on all SATAI dimensions:

- Total SATAI: $M_{use} = 3.57$ vs. $M_{no\ use} = 3.09$; $t(68.48) = 3.13$, $p = .003$, $d = 0.63$
- Cognitive: $M_{use} = 3.72$ vs. 3.31; $t(62.02) = 2.82$, $p = .007$, $d = 0.59$
- Affective: $M_{use} = 3.75$ vs. 3.22; $t(61.02) = 3.31$, $p = .002$, $d = 0.70$
- Behavioral: $M_{use} = 3.36$ vs. 2.90; $t(70.84) = 2.35$, $p = .021$, $d = 0.47$

These medium to large effects suggest that previous exposure to ChatGPT is associated with more favorable attitudes toward GenAI but does not systematically reduce AI anxiety.

Lastly, bivariate correlations were computed between total and subscale scores of SATAI and AIAS. GenAI attitudes and AI anxiety were essentially uncorrelated in this sample. For example, the correlation between total SATAI and total AIAS was $r = -.05$, $p = .60$; correlations between SATAI and the learning and Job Replacement subscales were similarly small and non-significant (all $|r| < .09$, all $p > .38$).

This pattern suggests that students can simultaneously hold positive attitudes toward GenAI and moderate levels of GenAI related anxiety, especially regarding job replacement. In other words, liking GenAI in school does not automatically imply feeling relaxed about its broader societal consequences.

Table 5. Gender differences in SATAI and AIAS scale scores (N=109)

| **Measure** | **Gender** | **Mean** | **Std. Deviation** | **t(df)** | **p** | **d** | **partial $\eta^2$** |
|---|---|---|---|---|---|---|---|
| AIAS total | Male | 3.58 | 1.37 | 0.82 (87.45) | 0.416 | 0.16 | 0.008 |
| | Female | 3.36 | 1.27 | | | | |
| AIAS learning | Male | 3.18 | 1.46 | 0.50 (86.33) | 0.616 | 0.1 | 0.003 |
| | Female | 3.04 | 1.32 | | | | |
| AIAS Job Replacement | Male | 4.28 | 1.6 | 1.07 (96.05) | 0.286 | 0.21 | 0.012 |
| | Female | 3.93 | 1.7 | | | | |
| SATAI total | Male | 3.06 | 0.71 | −1.99 (101.12) | 0.049 | −0.38 | 0.038 |
| | Female | 3.35 | 0.83 | | | | |
| SATAI cognitive | Male | 3.09 | 0.64 | −4.71 (95.23) | <.001 | −0.91 | 0.189 |
| | Female | 3.68 | 0.67 | | | | |
| SATAI affective | Male | 3.23 | 0.71 | −1.70 (101.84) | 0.093 | −0.32 | 0.027 |
| | Female | 3.49 | 0.84 | | | | |
| SATAI behavioral | Male | 2.91 | 0.92 | −1.21 (100.85) | 0.23 | −0.23 | 0.014 |
| | Female | 3.14 | 1.06 | | | | |

### 4.3 Qualitative results

The semi-structured interviews with 36 students revealed four pedagogical affordances and three constraints associated with the use of ChatGPT-5 during the intervention, as presented in Table 6.

#### 4.3.1 Pedagogical affordances

**Generating new knowledge (N=29)**

The first affordance relates to generating new knowledge. The majority of the students (N=29) viewed ChatGPT-5 as a tool for expanding their knowledge on a specific topic. Moreover, after realizing during the intervention that ChatGPT-5 could make errors, the students opted asking questions on topic they had prior knowledge as a way of feeling on top of the response in case it hallucinates. A small subset of the sample (N=7) still considered that ChatGPT-5 would provide new knowledge for them, without questioning the generated output, highlighting the need for critical evaluation.

**Immediate feedback (N=18)**

Immediate feedback was the second affordance that was identified by the sample (N=18) commented on how the response time is a key pedagogical advantage. Student responses were split into three categories: (a) they commented on the speed while requesting feedback on their own texts/writing, (b) when providing summaries of videos and attached files and lastly (c) the instant responses help keep the students engaged, who would otherwise be discouraged to use it on a daily basis. This rapid feedback is perceived as motivating as well as time-efficient for the students.

**Friendly and Familiar user interface (N=16)**

The third affordance deals with the friendly user interface. A significant number of students (N=16) commented how they felt the user interface felt "familiar" to social media, as well as intuitive.

**Skill development (N=10)**

The fourth Affordance that was identified has to do with Skill development. 10 students perceived that ChatGPT-5 helped them promote a number of skills including computational thinking (N=8), problem-solving (N=6), critical thinking (N=4) as well as digital literacy (N=2).

### 4.2.2 Constraints

**Uncertainty about content accuracy (N=21)**

The first one deals with uncertainty regarding the accuracy of the content generated. Moreover, (N= 21) expressed concerns regarding the truthfulness and the quality of ChatGPT-5 responses. Some students (N=9) commented on the need for generated content verification, which could be time-consuming. When questioned regarding reading the bottom of ChatGPT5's prompt box that stated that "ChatGPT may reply wrongfully", all 21 of the students read it, but nearly all of them (N=20) suggested it may have to do with mathematical calculations rather than provide incorrect data during hallucination.

**Anxiety related to AI feedback (N=11)**

The second constraint deals with anxiety concerns. 11 students reported anxiety while interacting with ChatGPT-5. Mostly every response had to do with ChatGPT-5's ability to provide feedback on the students input, with a continuous loop of suggestions for revisions, that made the students feel that their work would never be "good enough".

**Privacy and Data protection Concerns (N=4)**

The third and last constraint has to do with Privacy concerns. Four (N=4) students raised privacy and data protection concerns, emphasizing their uncertainty as to where the information they type is stored and who can access it. Moreover, some of them were reluctant to provide personal information when registering their account as well as during the intervention that required personal information for a personalized resume.

Table 6. Pedagogical affordances and constraints of using ChatGPT-5 in education (N=36)

| **Pedagogical affordances for teaching and learning** | **N** | **Limitations and issues of use** | **N** |
|---|---|---|---|
| Generating new knowledge (Expanding prior understanding) | 29 | Uncertainty regarding content accuracy and reliability | 21 |
| Immediate feedback | 18 | Anxiety related to AI feedback | 11 |
| Friendly and intuitive user interface | 16 | Privacy and data-protection concerns | 4 |
| Skill development (Computational thinking, problem-solving, critical thinking, digital literacy) | 10 | - | - |

## 5 Discussion

The purpose of this mixed-method study was to examine secondary students' acceptance, perceptions, affordances, anxiety, and constraints regarding the use of ChatGPT-5 in the Greek curriculum course "Technology". The intervention took place across three days and incorporated both structured activities and open exploration. Quantitative data was collected using the SATAI and AIAS scales, while qualitative data was obtained through semi-structured interviews. Overall, the findings reveal moderately positive attitudes toward GenAI, combined with low-to-moderate levels of learning-related anxiety and more pronounced concerns about job replacement.

Regarding RQ1 which examined students' attitudes, the SATAI results indicate that students hold moderately positive attitudes towards GenAI in education rather than strong enthusiasm about its use. Students viewed learning about GenAI in school as relevant and beneficial for the future. However, students' behavioral intention scores were consistently lower, especially when discussing AI-related careers or engaging in extracurricular AI-related activities, are items that received the lowest mean scores. This gap between attitude and behavior has been well documented in several approaches regarding technology adoption (e.g., Ajzen, 1991; Davis, 1989) and more recent emerging trends in education (Sivenas & Koutromanos, 2022b). More recent studies in the field of GenAI, identified a similar pattern where young learners do realize the use case of AI in education, but they do not necessarily believe they will use It in the future (Akgun & Greenhow, 2021; Kohnke et al., 2023). In the current study, prior experience with ChatGPT was clearly associated with more positive attitudes on all SATAI dimensions, without, however, systematically reducing anxiety levels, suggesting that familiarity with GenAI supports acceptance but does not automatically alleviate students' concerns.

In relation to RQ2 which examined student AI anxiety, the AIAS scale revealed overall moderate anxiety levels, with a clear distinction between learning related concerns and job replacement concerns. Learning-related items were generally close to or slightly below the midpoint of the scale, indicating modest unease about understanding and keeping up with AI techniques and products. In contrast, Job Replacement items were clearly higher, reflecting students' worries that GenAI may eventually outpace human skills and threaten future employment opportunities. This in turn aligns with similar studies in research literature where students believe that GenAI will outpace human skills and thus threaten future employment opportunities (Zhang & Dafoe, 2019). At the same time, attitudes toward GenAI and AI anxiety scores were essentially uncorrelated, indicating that students can hold positive views of GenAI in school while simultaneously being worried about its broader impact on work and society. These findings are further enhanced by the qualitative results. Moreover, the students expressed their concerns about the accuracy of the ChatGPT-5 output, a finding that is consistent with similar work on GenAI hallucinations (Valeri et al., 2024; Yim & Su, 2024). Additionally, the students reported stress on the continuous feedback loop, where GenAI kept suggesting text revisions, thus implying that their work will never be "good enough". This finding shows how AI-driven feedback can evoke unsettling emotions and the feeling of evaluation pressure when learners perceive the system to be either unpredictable or overly critical (Deepshikha, 2025; Rapp et al., 2025). Lastly, the job-replacement dimensions, align with emerging global trends, where students believe that GenAI may outpace them and thus reduce future employment opportunities (Zhang & Dafoe, 2019). This study contributes to a growing body of evidence that anxiety is a psychological factor that influences GenAI adoption in teaching and learning, especially in pre-university students.

Addressing RQ3 and RQ4 the qualitative data revealed four affordances of ChatGPT-5, namely: generating new knowledge, receiving immediate feedback, interacting through a familiar interface, and developing specific cognitive skills. Similar findings were identified in research literature (Bitzenbauer, 2023; Cheng & Chang, 2024) with an emphasis on real-time feedback, that was identified as strong predictor of learning gains (Ng et al., 2024). A particular clarification is that students often used ChatGPT-5 in order to explain content that they already partially understood, frequently asking to "explain it like I'm seven years old", or to provide simpler analogies in general. This finding highlights how students did not use GenAI to replace instruction, but rather to deepen and clarify knowledge and concepts they already possessed. Similar patterns where students primarily use GenAI in order to receive clarifications or rephrase concepts rather than obtain answers from scratch have been observed in educational settings (Bitzenbauer, 2023; Valeri et al., 2024).

A central qualitative finding concerns how students adapted their use of ChatGPT-5 after experiencing hallucinations. Many reported that, in future, they would primarily consult the chatbot on topics they already "knew something about," so that they could judge whether its answers were correct and ignore misleading ones. This pattern is conceptualized as epistemic safeguarding: a student-driven strategy in which GenAI is deliberately confined to "safe knowledge zones" where prior understanding can be used to monitor and correct the system's outputs. This differs from classic notions of trust calibration in human-AI interaction, where trust is adjusted in response to perceived system reliability (Lee & See, 2004; Okamura, 2020). Here, students were not simply learning when to trust the system; they were constraining where to use it in order to preserve epistemic control and minimize the risk of being misled. Epistemic safeguarding goes beyond generic advice to "fact-check AI outputs" (Long & Magerko, 2020; Yim & Su, 2024) and points to a more proactive, metacognitive form of critical AI literacy in which adolescents actively structure their own interaction boundaries with GenAI.

Moreover, during the intervention students were encouraged to identify and discuss hallucination instances openly, which appears to have strengthened their critical GenAI literacy. Rather than disengaging from AI or over-relying on it, students positioned themselves as evaluators, using GenAI to clarify familiar content while avoiding dependence on it in unfamiliar domains. This finding extends work on critical AI literacy (Long & Magerko, 2020; Klar, 2025) by demonstrating that exposure to AI limitations can promote epistemic agency in adolescents.

Lastly, three constraints were identified during the semi-structured interviews that deal with uncertainty about accuracy, anxiety as well as privacy concerns. Even though similar findings were identified in previous works (Valeri et al., 2024; Yim & Su, 2024), the current study builds on the introduction of hallucination and the inaccuracy of generated output, that may have caused increased anxiety to students, that may have used GenAI in the past and considered its responses as valid without the need of verification. The constraints that were identified highlight how GenAI should be accompanied among others by transparency and ethical use protocols (Yim & Su, 2024).

Overall, this study demonstrates that while secondary grade students recognize the educational value of GenAI and generally hold moderately positive attitudes toward its use, they also experience non-trivial levels of anxiety, particularly around job replacement, and are not willing to accept AI-generated content uncritically. The identification of epistemic safeguarding as a student-driven strategy for managing AI hallucinations represents a novel contribution to the GenAI-in-education field and has significant implications for the design of AI literacy interventions in K-12 settings, where fostering critical AI literacy and epistemic agency becomes as important as teaching technical skills.

From a computational intelligence standpoint, epistemic safeguarding can be interpreted as a form of human-in-the-loop control over a non-deterministic generative model. Rather than passively accepting model outputs, students dynamically regulate when and where to query the system, using their prior knowledge as an external constraint on the search space of possible AI responses. This kind of adaptive human-AI interaction is crucial for integrating powerful but fallible computational intelligence systems into everyday learning practices.

## 6 Limitations and future directions.

While this study provides valuable insights on the use of GenAI in Greek secondary level students, a number of limitations were identified, opening pathways for future research.

The first limitation deals with the sample. Moreover, this study used a convenient sample of 109 students from a specific geographical and cultural context from Greece. This limits the generalizability of the findings. Future studies could use a larger randomized sample across diverse countries and educational systems in order to examine further the cross-cultural applicability of epistemic safeguarding that was identified within this study. Comparative studies in different regions (e.g., Europe, Asia, America) could provide insights as to how cultural attitudes towards technology and education may influence GenAI adoption.

The second limitation deals with time constraints. Even though the intervention was split into three days, the students had 485 minutes (~ 8 hours) in order to interact with ChatGPT-5. Even though there is no minimal amount of time to consider for the novelty effect, it is a limitation that should be considered. Future studies should implement longitudinal designs tracking students over a semester of a full academic year. This would provide valuable insights into whether epistemic safeguarding is a novice strategy or in fact a stable component of digital literacy. It would also offer additional insights into GenAI's impact on the students' anxiety levels.

Thirdly, the exploration on the novel concept of epistemic safeguarding is in its preliminary state. While the data presented within this work illustrates this behavior, its psychological impact and learning outcomes are not yet fully measured. Future research could develop a quantitative scale in order to measure attitudes towards epistemic safeguarding. Additionally, more studies are required in order to identify what triggers epistemic safeguarding when compared to fact-checking that has been identified in previous research literature (Long & Magerko, 2020; Yim & Su, 2024).

The fourth limitation deals with the type of intervention that was utilized. As it was a text-based approach the students did not experience the ability to talk with the GenAI and completely avoided voice-based conversations. Future research could focus in voice-based interventions, or a combination of text and voice in order to examine further how these strategies influence attitudes, anxiety and trust towards GenAI.

Finally, this study focused on student samples. For a successful introduction of technology in the classroom more than one stakeholder is involved, including teachers, administrators as well as parents. Future research could adopt a multi-stakeholder approach in order to develop a deeper, holistic understanding of the requirements for a pedagogical framework for the successful use of GenAI in secondary education.

Overall, the findings highlight that the educational value of GenAI does not only depend on advances in computational intelligence architectures, but equally on how students are supported to develop critical, hallucination-aware interaction strategies. Designing curricula that explicitly address both dimensions (the underlying computational intelligence and the human practices around it) may be key for sustainable GenAI integration in secondary education.

**Funding**

This research received no external funding

**Institutional Review Board Statement**

According to National regulations and institutional policies for research in Greek schools, formal ethics board approval was not required for this study, as it took place during a structured course within the National curriculum. Permission to conduct the research was granted by the principals of the participating schools. Written informed consent was obtained from parents/guardians, and student assent was obtained prior to participation.

**Informed Consent Statement**

Written informed consent was obtained from all parents/guardians of participating students. Student assent was obtained prior to participation.

**Data Availability Statement**

Data will be made available upon request.

**Conflict of Interest**

The author declares no conflict of interest.